\documentclass{article}

\newif\ificassppaperkit
\IfFileExists{spconf.sty}{%
  \icassppaperkittrue
  \usepackage{spconf}%
}{%
  \icassppaperkitfalse
  \usepackage[
    letterpaper,
    top=25mm,
    bottom=25mm,
    left=19mm,
    right=19mm,
    columnsep=6mm
  ]{geometry}%
  \twocolumn
  \pagestyle{empty}%
}

\usepackage{amsmath,amssymb}
\usepackage{booktabs}
\usepackage{graphicx}
\usepackage{multirow}
\usepackage{array}
\usepackage{url}
\usepackage[table]{xcolor}
\usepackage{makecell}
\usepackage{etoolbox}

\makeatletter
\ificassppaperkit
  \def\name#1{\gdef\@name{{\em #1}}}
\else
  \newcommand{\name}[1]{\author{#1}}
  \newcommand{\address}[1]{\date{\small #1}}
  \newenvironment{keywords}
    {\par\small\noindent\textbf{Index Terms---}\ignorespaces}
    {\par}
\fi
\makeatother

\definecolor{bettercell}{RGB}{235,248,239}
\definecolor{worsecell}{RGB}{253,238,238}
\definecolor{normalcell}{RGB}{232,232,232}

\title{Joint Residual Reweighting for Classifier-Free Guidance\\
in Flow-Matching Zero-Shot TTS}

\name{\shortstack[c]{
Runwu Shi$^{1}$,
Yujin Wang$^{2}$,
Hongjin Song$^{3}$,
Jiang Wang$^{1}$,
Yaozhong Kang$^{1}$,
Nabeela Khan$^{1}$,
\\[-0.2em]
Weiqiao Shan$^{4}$,
Benjamin Yen$^{1}$,
Ashizawa Takeshi$^{1}$,
Chunxiang Jin$^{5}$,
Kazuhiro Nakadai$^{1}$
}}

\address{
$^{1}$Institute of Science Tokyo,
$^{2}$Wuhan University,
$^{3}$BIT,
$^{4}$Northeastern University,
$^{5}$Ant Group
}

\begin{document}

\maketitle

\begin{abstract}
Classifier-free guidance (CFG) is widely used in flow-matching-based
zero-shot text-to-speech (TTS), where generation is conditioned on text
content and a speech prompt. Standard CFG uses a single guidance weight
for their joint conditional effect, while branch-selective guidance
emphasizes text or speaker conditioning and can introduce a trade-off
between text accuracy and speaker similarity. In this paper, we revisit
CFG under independently masked conditions and decompose the guidance
field into text, speaker, and joint residuals. We show that
condition-specific branch differences couple the joint residual with
the corresponding text or speaker residual under a shared weight.
Trajectory analysis further shows that the joint residual varies over
flow time and contains information that cannot be represented by
reweighting the text and speaker residuals alone. Based on these
observations, we propose joint residual reweighting, which assigns
independent weights to the three residuals. Experiments on F5-TTS, CosyVoice2, and GLM-TTS across three evaluation sets show overall improvements in speaker similarity and text accuracy over the default CFG settings without retraining.
\end{abstract}


\section{Introduction}

Zero-shot text-to-speech (TTS) aims to synthesize speech for an unseen
speaker from a short reference utterance. The generated speech should
follow the target text while preserving the reference speaker's voice.
Recent flow-matching-based TTS systems model speech generation through a conditional flow-matching process with text and reference speech, and have shown strong zero-shot performance \cite{chen2025f5tts,eskimez2024e2,liu2026cross,wang2025eftts}.

Classifier-free guidance (CFG) is widely used in these models to control
the strength of conditional information during inference. Standard CFG
strengthens the joint effect of text and speaker conditioning, but does
not provide separate control over the two. Recent studies have therefore
explored condition-specific guidance by constructing additional branches
with one condition masked and using branch differences as guidance
directions \cite{li2024voiceldm,yang2024dualspeech,zheng2025selectivecfg}.
These approaches provide finer control, but also expose a trade-off
between text accuracy and speaker similarity: stronger
speaker-oriented guidance can improve speaker similarity while
degrading text accuracy \cite{zheng2025selectivecfg}.

A key limitation of this branch-difference view is that each difference
is treated as a single direction. Our analysis suggests that this trade-off is related to the fact that the effects of the two conditions are not fully independent: a
speaker-oriented branch can also affect text-related information, and
vice versa. We hypothesize that this cross-condition interference is
caused by a component that emerges only when text and speaker conditions
are jointly present. This interaction can be viewed as a \emph{joint
residual} component, which is not explicitly modeled by existing
condition-specific guidance. By considering the four branches obtained
by independently masking the two conditions, we isolate this joint
residual and find that it is non-negligible, varies over flow time, and
cannot in general be represented by rescaling the text and speaker
residuals.

Based on this observation, we introduce \textbf{joint residual
reweighting}, which explicitly decomposes the guidance field into text,
speaker, and joint residuals and reweights them independently. This
provides direct control over the joint component without retraining or
additional solver steps. Experiments on F5-TTS \cite{chen2025f5tts}, CosyVoice2 \cite{du2024cosyvoice2}, and GLM-TTS \cite{cui2025glm}
with LibriSpeech-PC, SEED-EN, and SEED-ZH show improvements in both
content accuracy and speaker similarity over default CFG settings.
Applying the method only at selected flow steps further retains these
gains with fewer additional branch evaluations.


\begin{figure*}[t]
  \centering
  \includegraphics[width=0.9\textwidth]{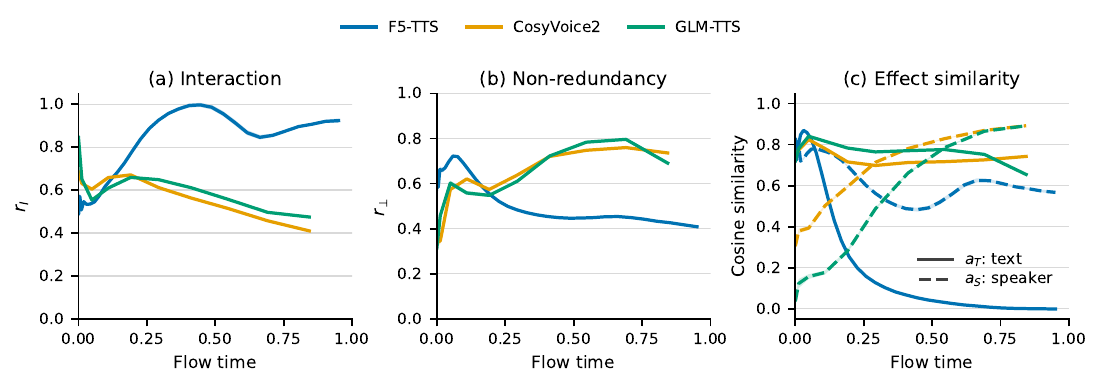}
\vspace{-1.0em}
\caption{Interaction dynamics along CFG trajectories, averaged over
500 LibriSpeech-PC test-clean samples. (a) Relative interaction
magnitude $r_I$. (b) Energy outside $\mathrm{span}(T,S)$. (c) Text and speaker
effect similarities $a_T$ (solid) and $a_S$ (dashed).}
  \label{fig:case-study}
\end{figure*}

\vspace{-0.5em}
\section{Method}
\vspace{-0.5em}

\subsection{Text--speaker residual factorization}
We consider a flow-matching TTS model with velocity predictor $f$. At a given
sampling state $x_t$ and flow time $t$, the generator is conditioned on content
information $c_{\mathrm{text}}$ and speaker information $c_{\mathrm{spk}}$.
Keeping $(x_t,t)$ fixed and independently masking the two conditions gives four velocity predictions:
\vspace{-1.0em}
\setlength{\jot}{2pt}
\begin{align}
v_{\mathrm{null}}
&=f(x_t,t,\varnothing_{\mathrm{spk}},
\varnothing_{\mathrm{text}}), \notag\\
v_{\mathrm{text}}
&=f(x_t,t,\varnothing_{\mathrm{spk}},c_{\mathrm{text}}), \notag\\
v_{\mathrm{spk}}
&=f(x_t,t,c_{\mathrm{spk}},\varnothing_{\mathrm{text}}), \notag\\
v_{\mathrm{full}}
&=f(x_t,t,c_{\mathrm{spk}},c_{\mathrm{text}}).
\label{eq:branches}
\end{align}
They denote the unconditional, text-only, speaker-only, and fully conditioned
predictions, respectively.

Taking $v_{\mathrm{null}}$ as a common reference, the isolated text and speaker
effects are
\begin{equation}
T=v_{\mathrm{text}}-v_{\mathrm{null}},\qquad
S=v_{\mathrm{spk}}-v_{\mathrm{null}}.
\label{eq:main_effects}
\end{equation}
Here, $T$ measures the response to adding the text condition without speaker
conditioning, while $S$ measures the response to adding the speaker condition
without text conditioning. These differences correspond to the null-based
condition directions used in separated multi-condition CFG.

If text and speaker conditioning were purely additive, the fully conditioned
prediction would satisfy
\begin{equation}
v_{\mathrm{full}}=v_{\mathrm{null}}+T+S.
\end{equation}
Joint conditioning may, however, produce a response beyond this additive
combination. We define the deviation as
\begin{align}
I
&=(v_{\mathrm{full}}-v_{\mathrm{null}})-T-S \notag\\
&=v_{\mathrm{full}}-v_{\mathrm{text}}-v_{\mathrm{spk}}
+v_{\mathrm{null}}.
\label{eq:interaction}
\end{align}
We refer to $I$ as the joint residual, which captures the deviation of joint
text--speaker conditioning from the additive combination of the two isolated
effects. The fully conditioned prediction can therefore be written as
\begin{equation}
v_{\mathrm{full}}
=v_{\mathrm{null}}+T+S+I.
\label{eq:factorization}
\end{equation}

This factorization also reveals how the joint residual enters
condition-specific guidance:
\begin{equation}
v_{\mathrm{full}}-v_{\mathrm{text}}=S+I,\qquad
v_{\mathrm{full}}-v_{\mathrm{spk}}=T+I.
\label{eq:contextual_effects}
\end{equation}
Thus, a speaker-oriented branch difference couples $S$ with $I$, while
a text-oriented difference couples $T$ with $I$. A shared guidance
weight adjusts the two components together. This motivates us to
examine whether $I$ is substantial and whether it can be represented
by reweighting $T$ and $S$ alone.


\vspace{-0.5em}
\subsection{Properties of the text--speaker interaction}
We examine the joint residual $I$ along the default CFG trajectories,
without applying our reweighting. We focus on three questions: whether
$I$ is substantial, whether it can be represented by reweighting $T$
and $S$, and how it modifies the two condition effects.

\textbf{Interaction magnitude.}
We first measure the magnitude of $I$ relative to the full--null
conditional direction:
{
\setlength{\abovedisplayskip}{4pt}
\setlength{\belowdisplayskip}{4pt}
\setlength{\abovedisplayshortskip}{2pt}
\setlength{\belowdisplayshortskip}{2pt}
\begin{equation}
r_I=
\frac{\lVert I\rVert_2}
{\lVert v_{\mathrm{full}}-v_{\mathrm{null}}\rVert_2}.
\label{eq:interaction_strength}
\end{equation}
}
A larger $r_I$ indicates a stronger deviation from additive
conditioning. Figure~\ref{fig:case-study}(a) shows a substantial joint
residual in all three systems, with distinct dynamics over flow time.
F5-TTS approaches one around the middle of sampling and remains high,
whereas CosyVoice2 and GLM-TTS gradually decrease in later stages.

\textbf{Non-redundancy.}
A substantial $I$ does not necessarily require an independent guidance
component, since it may still be represented by reweighting $T$ and
$S$. We therefore measure its energy outside $\mathrm{span}(T,S)$.
After flattening the velocity tensors over frames, we
obtain the least-squares projection
$I_{\parallel}=a^{\star}T+b^{\star}S$ and define
{
\setlength{\abovedisplayskip}{4pt}
\setlength{\belowdisplayskip}{4pt}
\begin{equation}
r_{\perp}=
\frac{\lVert I-I_{\parallel}\rVert_2^2}
{\lVert I\rVert_2^2}.
\label{eq:orthogonal}
\end{equation}
}
Figure~\ref{fig:case-study}(b) shows substantial energy outside this
span for all three systems. Notably, CosyVoice2 and GLM-TTS exhibit
higher $r_{\perp}$ than F5-TTS in the middle and later stages despite
their smaller $r_I$. Thus, the magnitude and non-redundancy of the
joint residual capture different properties.

\textbf{Effect modification.}
Cosine similarity has been used to examine directional alignment along
CFG trajectories~\cite{castillo2025adaptive}. We examine how $I$
changes the text and speaker directions:
{
\setlength{\abovedisplayskip}{4pt}
\setlength{\belowdisplayskip}{4pt}
\begin{equation}
a_T=\cos(T,T+I),\qquad
a_S=\cos(S,S+I).
\label{eq:effect_similarity}
\end{equation}
}
Values close to one indicate little directional change.
Figure~\ref{fig:case-study}(c) shows distinct effects across models
and flow time. For F5-TTS, $I$ strongly alters the text direction in
later stages while its effect on the speaker direction is smaller.
CosyVoice2 and GLM-TTS show different profiles, with $a_S$ increasing
over time and eventually exceeding $a_T$.

Together, these results show that the joint residual is substantial,
not fully represented by $T$ and $S$, and affects the two condition
directions differently across models and flow time. This motivates
treating $I$ as an independent guidance component. These observations
also guide the model-specific coefficient selection in Sec.~3.1.

\vspace{-0.5em}
\subsection{Joint residual reweighting for CFG}

The preceding analysis motivates treating $I$ as an independent guidance
component. Under the factorization in Eq.~\eqref{eq:factorization},
standard full--null CFG can be written as
\begin{align}
v_{\mathrm{CFG}}
&=v_{\mathrm{full}}
+r(v_{\mathrm{full}}-v_{\mathrm{null}}) \notag\\
&=v_{\mathrm{full}}+r(T+S+I),
\label{eq:cfg_factorized}
\end{align}
where a single guidance weight jointly scales all three residuals.

We propose joint residual reweighting, which introduces an independent
guidance weight for each residual:
\begin{equation}
\begin{aligned}
v_{\mathrm{guided}}
={}&v_{\mathrm{full}}+g_TT+g_SS+g_II.
\end{aligned}
\label{eq:ours}
\end{equation}
Here, $g_T$, $g_S$, and $g_I$ control the additional text, speaker, and
joint guidance, respectively. Setting $g_T=g_S=g_I=r$ recovers standard
full--null CFG.

Standard CFG uses one guidance degree of freedom by jointly scaling
$T$, $S$, and $I$. Condition-specific guidance introduces two degrees of
freedom, but still couples $I$ with either $T$ or $S$. In contrast, our
formulation independently controls all three residuals through
$g_T$, $g_S$, and $g_I$.

The four prediction branches share the same $(x_t,t)$ and can be
evaluated in one batch. Our method requires no retraining or additional
ODE solver steps; the additional computation comes from evaluating
the extra condition branches at guided steps. We discuss reducing this
additional computation in Sec.~3.3.


\begin{table}[t]
\centering
\footnotesize
\setlength{\tabcolsep}{3.0pt}
\renewcommand{\arraystretch}{1.0}
\caption{Trajectory statistics and guidance weights.}
\label{tab:selection}
\begin{tabular}{@{}lccccccc@{}}
\toprule
Backbone
& $r$ (CFG)
& $\bar r_I$
& $\bar r_{\perp}$
& $\Delta_{\mathrm{align}}$
& $g_T$
& $g_S$
& $g_I$ \\
\midrule
F5-TTS
& 2.0
& 0.852 & 0.495 & -0.439
& 2.00 & 3.00 & 4.50 \\
CosyVoice2
& 0.7
& 0.547 & 0.673 & +0.008
& 0.70 & 1.20 & 0.95 \\
GLM-TTS
& 0.7
& 0.574 & 0.671 & +0.169
& 0.95 & 0.70 & 0.80 \\
\bottomrule
\end{tabular}
\vspace{-1.0em}
\end{table}

\begin{table*}[!t]
\centering
\footnotesize
\setlength{\tabcolsep}{6.0pt}
\renewcommand{\arraystretch}{1.08}

\caption{Results on three complete evaluation sets. Green and red cells denote
improvements and degradations over the best reproduced CFG result for the same
backbone, set, and metric, respectively. \textbf{\underline{Best}} results are
bold and underlined; \textbf{second-best} results are bold. Higher SIM and
lower WER/CER are better. $^{\dagger}$ Values are quoted from Selective
CFG~\cite{zheng2025selectivecfg}; $^{*}$ denotes our reproduction.}
\label{tab:main}

\begin{tabular}{llc cc cc cc}
\toprule
\multirow{2}{*}{Backbone}
& \multirow{2}{*}{Method}
& \multirow{2}{*}{Branches}
& \multicolumn{2}{c}{LibriSpeech-PC}
& \multicolumn{2}{c}{SEED-EN}
& \multicolumn{2}{c}{SEED-ZH} \\
\cmidrule(lr){4-5}
\cmidrule(lr){6-7}
\cmidrule(lr){8-9}
&&&
SIM $\uparrow$ & WER $\downarrow$
& SIM $\uparrow$ & WER $\downarrow$
& SIM $\uparrow$ & CER $\downarrow$ \\
\midrule

\multirow{4}{*}{GLM-TTS}
& CFG$^{*}$ (0.7, official) & 2
& 0.6673 & 0.0273
& 0.6712 & 0.0210
& \textbf{0.7618} & 0.0099 \\

& CFG$^{*}$ (1.0) & 2
& 0.6670 & \textbf{0.0272}
& 0.6739 &  \textbf{0.0207}
& 0.7612 & \textbf{0.0099} \\

& Selective CFG$^{*}$~\cite{zheng2025selectivecfg} & 2
& \cellcolor{bettercell}\textbf{0.6674}
& \cellcolor{worsecell}0.0279
& \cellcolor{bettercell}\textbf{0.6764}
& \cellcolor{normalcell}\textbf{0.0207}
& \cellcolor{worsecell}0.7608
& \cellcolor{worsecell}0.0100 \\

\cmidrule(lr){2-9}

& \textbf{Ours} & 4
& \cellcolor{bettercell}\textbf{\underline{0.6707}}
& \cellcolor{bettercell}\textbf{\underline{0.0270}}
& \cellcolor{bettercell}\textbf{\underline{0.6771}}
& \cellcolor{bettercell}\textbf{\underline{0.0198}}
& \cellcolor{bettercell}\textbf{\underline{0.7645}}
& \cellcolor{bettercell}\textbf{\underline{0.0090}} \\

\midrule

\multirow{4}{*}{F5-TTS}
& CFG$^{*}$ (1.5) & 2
& 0.6644 & 0.0210
& 0.6717 & 0.0153
& 0.7571 & 0.0168 \\

& CFG$^{*}$ (2.0, official) & 2
& 0.6745 & \textbf{0.0197}
& 0.6756 & \textbf{0.0143}
& 0.7618 & \textbf{0.0161} \\

& Selective CFG$^{\dagger}$~\cite{zheng2025selectivecfg} & 2
& \cellcolor{bettercell}\textbf{\underline{0.682}}
& \cellcolor{worsecell}0.022
& \cellcolor{bettercell}\textbf{\underline{0.690}}
& \cellcolor{worsecell}0.018
& \cellcolor{bettercell}\textbf{\underline{0.764}}
& \cellcolor{worsecell}0.019 \\

\cmidrule(lr){2-9}

& \textbf{Ours} & 4
& \cellcolor{bettercell}\textbf{0.6819}
& \cellcolor{bettercell}\textbf{\underline{0.0196}}
& \cellcolor{bettercell}\textbf{0.6779}
& \cellcolor{bettercell}\textbf{\underline{0.0136}}
& \cellcolor{bettercell}\textbf{0.7628}
& \cellcolor{bettercell}\textbf{\underline{0.0159}} \\

\midrule

\multirow{4}{*}{CosyVoice2}
& CFG$^{*}$ (0.7, official) & 2
& 0.6561 & \textbf{0.0212}
& 0.6603 & \textbf{0.0221}
& 0.7514 & 0.0138 \\

& CFG$^{*}$ (1.0) & 2
& 0.6585 & 0.0219
& 0.6634 & \textbf{0.0221}
& 0.7534 & \textbf{\underline{0.0130}} \\

& Selective CFG$^{\dagger}$~\cite{zheng2025selectivecfg} & 2
& \cellcolor{bettercell}\textbf{\underline{0.671}}
& \cellcolor{worsecell}0.025
& \cellcolor{bettercell}\textbf{0.666}
& \cellcolor{worsecell}0.026
& \cellcolor{bettercell}\textbf{\underline{0.763}}
& \cellcolor{worsecell}0.018 \\

\cmidrule(lr){2-9}

& \textbf{Ours} & 4
& \cellcolor{bettercell}\textbf{0.6690}
& \cellcolor{bettercell}\textbf{\underline{0.0211}}
& \cellcolor{bettercell}\textbf{\underline{0.6705}}
& \cellcolor{bettercell}\textbf{\underline{0.0219}}
& \cellcolor{bettercell}\textbf{0.7536}
& \cellcolor{worsecell}\textbf{0.0131} \\

\bottomrule
\end{tabular}

\vspace{-1.0em}
\end{table*}

\vspace{-1.0em}
\section{Experiments}
\subsection{Experimental setup and guidance configuration}

\hspace*{\parindent}\textbf{Backbones.}
We evaluate three open-source flow-matching zero-shot TTS systems with
different conditioning structures. F5-TTS~\cite{chen2025f5tts} directly
generates speech with a non-autoregressive flow matcher conditioned on
text and reference speech. CosyVoice2~\cite{du2024cosyvoice2} first
predicts semantic tokens with a language model, which are then passed
to a flow-based acoustic decoder together with speaker conditioning.
GLM-TTS~\cite{cui2025glm} similarly uses an autoregressive speech-token
model followed by a flow-based decoder, with speaker information
provided through its conditioning pathway. These systems therefore
cover both direct and cascaded flow-matching TTS architectures with
different forms of content and speaker conditioning.

\textbf{Evaluation.}
We evaluate on the complete LibriSpeech-PC (1,127 utterances), SEED-EN
(1,088), and SEED-ZH (2,020) sets. Content accuracy is measured by WER
for English and CER for Chinese, using Whisper large-v3~
\cite{radford2023robust} and Paraformer~\cite{gao2022paraformer},
respectively. Speaker similarity (SIM) is computed using WavLM speaker
embeddings~\cite{chen2022large}. F5-TTS uses 32 flow steps, while
CosyVoice2 and GLM-TTS use 10. Within each backbone, we keep the checkpoint, prompts, random seeds,
sampling schedule, and evaluation setup fixed across guidance settings.

\textbf{Guidance configuration.}
We use each backbone's default CFG sampler as the baseline and compute
the trajectory statistics in Sec.~2.2 on its unmodified trajectories.
Each statistic is averaged over flow time, weighted by solver intervals,
and then across utterances. We define
$\Delta_{\mathrm{align}}=\bar a_T-\bar a_S$ to characterize the relative
effect of the joint residual on the two condition directions, with negative
values indicating stronger modification of the text direction and positive
values stronger modification of the speaker direction.

Table~\ref{tab:selection} reports the trajectory statistics and final
guidance weights $g_T$, $g_S$, and $g_I$. We use
$\Delta_{\mathrm{align}}$ as an empirical heuristic for configuring
the residual weights. Since $I$ already contributes to the more strongly
modified direction, further increasing the corresponding main effect may
overemphasize it. We therefore retain its default CFG weight and adjust
the other condition weight together with $g_I$. Accordingly, F5-TTS
retains the default text weight, while GLM-TTS retains the default
speaker weight. For CosyVoice2, the alignment difference is near zero,
showing no clear preference; we retain the default text weight. The remaining
coefficients are selected through a small grid search for each backbone.

\vspace{-1em}
\subsection{Main results}

Table~\ref{tab:main} reports results on the three complete evaluation sets.
Relative to each backbone's released default CFG, the proposed reweighting
increases SIM and reduces WER/CER in all backbone--dataset combinations.
The relative benefits for speaker similarity and text accuracy vary across
models and evaluation sets, while the direction of improvement over the
default configurations is consistent.

For F5-TTS and GLM-TTS, the proposed method improves both metrics over
both evaluated CFG settings on all three sets. CosyVoice2 follows the
same pattern on the two English sets. On SEED-ZH, it achieves higher
SIM but slightly higher CER than the higher-scale CFG setting, while
still improving both metrics over the released default. This is the
only case where content accuracy does not improve over the best tested
standard CFG result. Overall, these comparisons suggest that
residual-specific reweighting offers a useful alternative to adjusting
a single global guidance scale.

Selective CFG~\cite{zheng2025selectivecfg} provides the closest comparison.
For GLM-TTS, the proposed method improves both metrics over our reproduced
selective setting on all three sets. The quoted F5-TTS and CosyVoice2
results generally combine higher SIM with higher WER/CER than the
proposed method, illustrating a different balance between speaker
preservation and text accuracy.

\begin{table}[t]
\centering
\footnotesize
\setlength{\tabcolsep}{3.0pt}
\renewcommand{\arraystretch}{1.0}
\caption{Weight ablation in residual coordinates on F5-TTS/LibriSpeech-PC.}
\label{tab:ablation_f5}
\begin{tabular}{@{}lccc@{}}
\toprule
Setting  & Guided velocity
& SIM $\uparrow$ & WER $\downarrow$ \\
\midrule

CFG
& $v_{\mathrm{full}}+2(T+S+I)$
& 0.6745
& 0.0197 \\

Equal $S/I$
& $v_{\mathrm{full}}+2T+3S+3I$
& \cellcolor{bettercell}0.6788
& \cellcolor{bettercell}0.0196 \\

Stronger $I$
& $v_{\mathrm{full}}+2T+3S+4.5I$
& \cellcolor{bettercell}\textbf{0.6819}
& \cellcolor{bettercell}0.0196 \\

$T/S$ control
& $v_{\mathrm{full}}+3T+3S+2I$
& \cellcolor{worsecell}0.6621
& \cellcolor{bettercell}\textbf{0.0180} \\

\bottomrule
\end{tabular}
\vspace{-1.0em}
\end{table}

\vspace{-1.0em}
\subsection{Ablation study and inference efficiency}
The proposed residual factorization suggests that the text, speaker, and
joint components play different roles in the speaker--text trade-off.
Table~\ref{tab:ablation_f5} reports an ablation on F5-TTS/LibriSpeech-PC,
with all settings expressed in the residual coordinates of
Eq.~\eqref{eq:ours}. The standard CFG sampler uses scale $r=2.0$,
corresponding to $(g_T,g_S,g_I)=(2.0,2.0,2.0)$. Increasing $g_S$ and
$g_I$ together improves speaker similarity while keeping WER nearly
unchanged, and further increasing $g_I$ yields an additional SIM gain.
In contrast, increasing the two isolated effects while leaving $g_I$
unchanged substantially reduces SIM despite a lower WER. These results
suggest that the joint residual has a distinct role in the guidance
trade-off and benefits from separate control.

Full-step guidance uses four branches rather than two in standard CFG,
increasing computation even when batched. We therefore apply the correction
only at selected steps and retain the released CFG elsewhere. Sparse schedules
are evaluated on the complete F5-TTS/LibriSpeech-PC set, with all settings
retaining 32 solver steps. In Table~\ref{tab:sparse}, \emph{Loc.} denotes the
corrected flow-step interval in start:end notation, with the end index excluded,
and \emph{Branch Evals.} counts the total condition-branch predictions across
all solver steps. Concentrating the correction in earlier steps is already
effective: six corrected steps improve both metrics with 76 branch evaluations,
reducing the additional evaluations over standard CFG from 64 to 12 (81.25\%).
Among the tested schedules, eight corrected steps yield the lowest WER, while
16 yield the highest SIM and match the baseline WER. Branch counts do not
directly measure runtime savings, and since the intervals are selected on the
same evaluation set, the results describe the observed trade-off rather than
held-out schedule generalization.

\begin{table}[t]
\centering
\footnotesize
\setlength{\tabcolsep}{2.0pt}
\renewcommand{\arraystretch}{1.0}
\caption{Sparse-step results on F5-TTS/LibriSpeech-PC.}
\label{tab:sparse}
\begin{tabular}{@{}lcccccc@{}}
\toprule
Setting & Total Steps & Guided Steps & Loc. & Branch Evals.
& SIM $\uparrow$ & WER $\downarrow$ \\
\midrule
CFG
& 32 & 0 & -- & 64
& 0.6745 & 0.0197 \\

Full
& 32 & 32 & $0{:}32$ & 128
& \cellcolor{bettercell}0.6819
& \cellcolor{bettercell}0.0196 \\

Sparse
& 32 & 6 & $1{:}7$ & 76
& \cellcolor{bettercell}0.6791
& \cellcolor{bettercell}0.0194 \\

Sparse
& 32 & 8 & $0{:}8$ & 80
& \cellcolor{bettercell}0.6794
& \cellcolor{bettercell}\textbf{0.0193} \\

Sparse
& 32 & 16 & $0{:}16$ & 96
& \cellcolor{bettercell}\textbf{0.6827}
& \cellcolor{normalcell}0.0197 \\
\bottomrule
\end{tabular}
\vspace{-1.0em}
\end{table}

\vspace{-1.0em}
\section{Conclusion}
\vspace{-1.0em}

Classifier-free guidance is widely used in flow-matching zero-shot TTS,
but condition-specific guidance can couple individual and joint effects.
We decompose the guidance field into text, speaker, and joint residuals,
and introduce joint residual reweighting for independent control.
Experiments across three TTS systems and three evaluation sets show
overall gains in both content accuracy and speaker similarity without
retraining. Sparse-step guidance retains these gains with fewer parallel branch evaluations.

\newpage

\IfFileExists{references.bib}{%
  \IfFileExists{IEEEbib.bst}{%
    \bibliographystyle{IEEEbib}%
  }{%
    \bibliographystyle{unsrt}%
  }
  \bibliography{references}
}{%
}

\end{document}